\documentclass[
aip,
reprint,
twocolumn,
amsmath,
amssymb
]{revtex4-1}

\usepackage[T1]{fontenc}
\usepackage[latin9]{inputenc}
\usepackage{eso-pic, rotating}
\usepackage{graphicx}
\usepackage{setspace}
\usepackage{amsmath}
\usepackage{enumerate}
\usepackage{textcomp}
\usepackage{amssymb}
\usepackage{relsize}
\usepackage{appendix}
\usepackage{babel}
\usepackage{amsfonts}

\usepackage{dcolumn}
\usepackage{bm}
\usepackage{mathptmx}

\AddToShipoutPicture{\put(580,170){\rotatebox{90}{\scalebox{0.9}{\color{lightgray}This article may be downloaded for personal use only. Any other use requires prior permission of the author and AIP Publishing.}}}}
\AddToShipoutPicture{\put(588,170){\rotatebox{90}{\scalebox{0.9}{\color{lightgray}This article appeared in G. S. Carmantini, F. S. Neves, M. Timme and S. Rodrigues, Chaos {\bf 31}, 093130 (2021)}}}}
\AddToShipoutPicture{\put(596,170){\rotatebox{90}{\scalebox{0.9}{\color{lightgray} and may be found at \url{https://doi.org/10.1063/5.0054485}.}}}}

\begin{document}

\preprint{AIP/123-QED} 

\title{Stochastic facilitation in heteroclinic communication channels}

\author{Giovanni Sirio Carmantini}
    \email{giovanni@carmantini.com}
\affiliation{foldAI, 81369 Munich, Germany}

\author{Fabio Schittler Neves}
    \email{fabio.neves@tu-dresden.de}
\affiliation{Chair for Network Dynamics, Center for Advancing Electronics Dresden (cfaed) and Institute for Theoretical Physics, TU Dresden, 01062 Dresden, Germany}

\author{Marc Timme}
    \email{marc.timme@tu-dresden.de}
\affiliation{Chair for Network Dynamics, Center for Advancing Electronics Dresden (cfaed) and Institute for Theoretical Physics, TU Dresden, 01062 Dresden, Germany}
\affiliation{Cluster of Excellence Physics of Life, TU Dresden, 01062 Dresden, Germany}
\affiliation{Lakeside Labs, 9020 Klagenfurt am W{\"o}rthersee, Austria}

\author{Serafim Rodrigues}
    \email{srodrigues@bcamath.org}
\affiliation{BCAM - Basque Center for Applied Mathematics, 48009 Bilbao, Bizkaia, Spain}
\affiliation{Ikerbasque, Basque Foundation for Science, 48013 Bilbao, Bizkaia, Spain}


\begin{abstract}
Biological neural systems encode and transmit information as patterns of activity tracing complex trajectories in high-dimensional state-spaces, inspiring alternative paradigms of information processing. Heteroclinic networks, naturally emerging in artificial neural systems, are networks of saddles in state-space that provide a transparent approach to generate complex trajectories via controlled switches among interconnected saddles. External signals induce specific switching sequences, thus dynamically encoding inputs as trajectories. Recent works have focused either on computational aspects of heteroclinic networks, i.e. \textit{Heteroclinic Computing}, or their stochastic properties under noise. Yet, how well such systems may transmit information remains an open question. Here we investigate the information transmission properties of heteroclinic networks, studying them as communication channels. Choosing a tractable but representative system exhibiting a heteroclinic network, we investigate the mutual information rate (MIR) between input signals and the resulting sequences of states as the level of noise varies. Intriguingly, MIR does not decrease monotonically with increasing noise. Intermediate noise levels indeed maximize the information transmission capacity by promoting an increased yet controlled exploration of the underlying network of states. Complementing standard stochastic resonance, these results highlight the constructive effect of stochastic facilitation (i.e. noise-enhanced information transfer) on heteroclinic communication channels and possibly on more general dynamical systems exhibiting complex trajectories in state-space.
\end{abstract}

\keywords{Spiking Neural Networks, Heteroclinic Networks, Coupled Oscillators, Stochastic Facilitation, Analog Computation}
\maketitle 

\begin{quotation}
In biological neural systems, computation and information transmission are not always independent processes, but rather a mixed process in which feedback connections between regions may surpass feed-forward ones in numbers, generating complex patterns of neural activity. This suggests that these systems not only transmit but simultaneously preprocess information. Remarkably, neural computing is robust despite intrinsically noisy environments (e.g. in the brain) and despite a lack of reproducibility of state space trajectories, such that spike pattern activity does not robustly repeat when performing the same given task. Heteroclinic networks, naturally emerging in a class of artificial neural networks, offer one efficient way to emulate concurrent neural information transmission and computation. In this paradigm, even small symmetrical systems can already express a large number of complex trajectories, as their number increases exponentially with system size. Moreover, heteroclinic information processing and transmission properties are deterministic and reproducible, but become stochastic if noise is present. We here study heteroclinic networks as noisy communication channels. Specifically, we measure the mutual information rate (MIR) between input signals and the resulting state-space trajectories -- the sequence of visited unstable state vicinities. We reveal a constructive effect of noise -- stochastic facilitation -- on the channel. Noise of small to moderate strength increases the MIR before decreasing it as the input signal is completely overridden. It may be possible to extend this result to different dynamical and neural systems exhibiting complex trajectories in state-space, as many may also rely on networks of unstable states to work.
\end{quotation}

\section{Introduction}

The brain makes use of a variety of energy-efficient and robust encoding strategies to process information. Examples vary from population coding, where information is encoded as statistical properties of neural populations, to precise spike timings, where information capacity is only bounded by noise \cite{bohte2004evidence, thorpe1990spike}. Moreover, studies of highly active neural systems, e.g. the vertebrate's olfactory bulb or the fly's antennal lobe \cite{laurent1994odorant}, suggest that information may also be encoded as patterns of activity tracing complex trajectories in high-dimensional state spaces. Together, such findings have inspired the conceptualization of a number of novel dynamical mechanisms for information processing, yielding insights on central interdisciplinary issues such as pattern generation \cite{Steingrube2010pattern, ijspeert2005simulation}, pre-processing to facilitate computations \cite{maass2002real,jaeger2004harnessing, Rabinovich2008processing, lukovsevivcius2009reservoir, Larger2012photonic}, noise-enhanced information storage \cite{kirst2016dynamic} and computing and signal encoding \cite{Ashwin2005, Mazor2005representation, Wordsworth2008, Neves2012, Neves2020Reconfigurable}.

\begin{figure}[]
\begin{centering}
\includegraphics[width=8.5cm,angle=0]{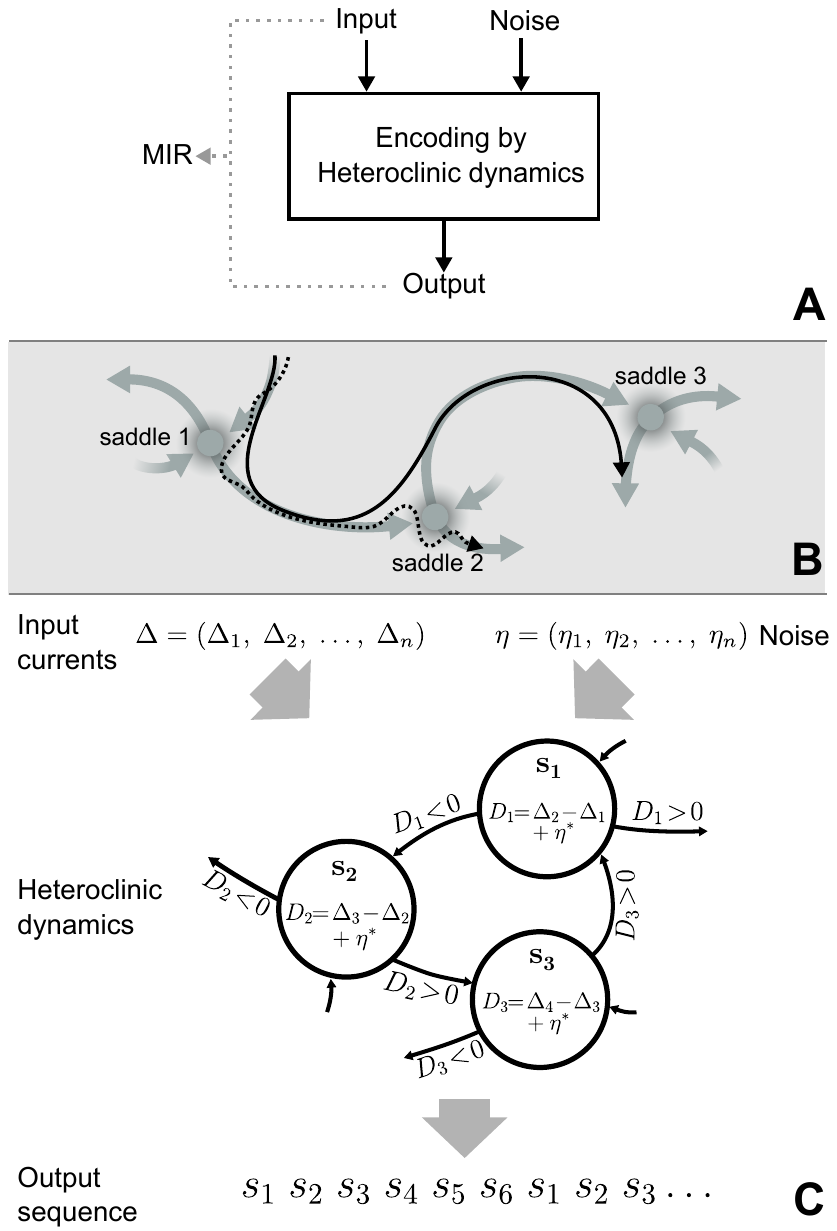}
\par\end{centering}
\caption[1st-entry]{ \small \textbf{Schematic of a heteroclinic network as an input-output noisy communication channel.} (\textbf{A}) The heteroclinic network encodes its input in the presence of noise, and produces an output; the mutual information rate (MIR) between input and output quantifies how much information the system is able to transmit. (\textbf{B}) The encoding process is based on the heteroclinic switching between saddles. In a noiseless system (black trajectory), the relative strengths of the components of the input signal completely determine the transition from one saddle to the next, by driving the state of the system towards a specific unstable direction and through an associated heteroclinic connection. When noise is present (dotted trajectory) and strong enough, it can overcome the deterministic effects of the input near a saddle and may push the dynamics towards another unstable direction, thus adding some randomness to the process. (\textbf{C}) A more detailed representation of a heteroclinic network seen as a currents-in, sequences-out noisy communication channel. Each saddle vicinity $s_i$ can be seen as a process comparing a pair of input components $\Delta_i$ under noisy conditions. The input signal causes the system to continuously switch between states, thus producing a sequence of visited states.}
  \label{fig:saddle_sequence}
\end{figure}

A particularly transparent example of dynamics supporting complex trajectories in high dimensional state-spaces are heteroclinic networks, which naturally emerge in systems of spiking \cite{Ernst1998delay, Timme2002prevalence, TWG2003} and non-spiking oscillators \cite{Krupa1997,Wordsworth2008}, as well as non-oscillatory systems \cite{bick2010occurrence} in artificial and, potentially, in natural systems. A heteroclinic network is a network composed of saddle states in which each connection is a heteroclinic orbit, i.e. an orbit connecting part of the unstable manifold of a saddle with the stable manifold of a second saddle\footnote{Referring to these saddles as ``states'' stems from a parallel with finite state machines. In fact, when describing a perturbation-driven sequence of saddle-to-saddle switching in heteroclinic networks, it is rather natural to regard saddles in the network as states, and heteroclinic connections between saddles as transitions, with the resulting sequence of visited states being a function of a starting state and applied input. For this reason, we here refer to saddles in a heteroclinic network interchangeably as ``saddle states'' or, simply, ``states''.}. 
Here, we study a particular class of symmetrical heteroclinic networks, in which each saddle is locally surrounded by basins of attraction of other saddles and these basins combined locally have full measure. Due to the pulse-coupled nature of the class of systems studied, almost all trajectories perturbed away from a given saddle periodic orbit would converge to one other saddle in finite time. All basin volume of each attractor periodic orbit is not in that local volume, but remotely located, close to other saddles in the network \cite{Ernst1995Synchronization,Ernst1998delay,Timme2002prevalence,TWG2003,AT2005_2,Neves2009}. Through sufficiently small perturbations applied to the system at one saddle state, state trajectories thus end up in other symmetry-related saddles. In this sense, the network of saddle states constitutes a "clean" heteroclinic network as termed by Field \cite{Field2016Patterns,Ashwin2020Almost,Podvigina2020Asymptotic}.

It has been shown that heteroclinic networks can support information encoding \cite{Ashwin2005, Wordsworth2008, Neves2012} via switching dynamics \cite{Krupa1997, TWG2003, AT2005_2, KGT2009, Neves2009} among its constituting states. Furthermore, through the \textit{Heteroclinic Computing} paradigm \cite{Neves2012}, such networks have been shown to be capable of implementing logic gates and operators, and thus support arbitrary n-ary computations. In this paradigm, switchings are driven by external signals serving as inputs, forcing the dynamics towards specific unstable directions \cite{AT2005,Neves2009} (see Figure \ref{fig:saddle_sequence}), thus generating a complex trajectory approaching saddles sequentially. At each switching event (i.e. near each saddle) the dominant signal components on the unstable manifolds are computed. Moreover, if an external signal persists for long enough, a cyclic sequence of states is established \cite{Wordsworth2008, Neves2012}, resulting in the computation of a partial rank order of the signal's components, also know as $k$-winners-take-all, where the $k$ strongest components out of a total of $N$ are identified.   

Alongside computation, information transmission is a fundamental function of neural systems, which is performed in an intrinsically noisy environment via complex patterns of activity. Here, to investigate the information transmission properties of bio-inspired dynamical systems exhibiting complex trajectories in state-space, we study heteroclinic networks as input-output noisy communication channels. Previous work has shown that 1) noise adds a stochastic component to an otherwise deterministic switching process and, thereby, modifies transition probabilities between saddles in a network-of-states \cite{Stone1999Noise, Armbruster2003Noisy, Bakhtin2010Small, Ashwin2010Designing, bakhtin2011noisy, ashwin2016quantifying,Voit2019Dynamical}, 2) perturbations at saddle states grow exponentially with time \cite{Neves2009}, and 3) noise also accelerates the saddle-to-saddle, defining an upper boundary for the switching time depending on the noise level itself \cite{neves2017noise}. To capture all these aspects in one measure, we characterize the heteroclinic channel by computing the mutual information rate (MIR) between external input signals and the resulting sequences of states, for varying noise level.

Typically, noise tends to lower the performance of communication channels, as it introduces errors in the transmission of information, thus reducing confidence in the reconstruction of the original signal. In contrast, we here report that intermediate levels of noise maximize the information transmission capacity of the system -- a stochastic facilitation effect. The mechanism underlying such effect relies on two factors: first, noise can reduce the time spent in the vicinity of each saddle (see \cite{neves2017noise}, as well as Supplementary Material); and second, it can promote an increased yet controlled exploration of the underling network of states. As a consequence, the MIR between input signals and the sequences of visited states depends non-monotonically on the noise levels. The MIR increases for intermediate levels of noise, before monotonically decreasing, until the resulting switching direction at each saddle becomes virtually random. 

We argue that our results are general to systems exhibiting heteroclinic networks, because they arise simply from the stochastic nature of the dynamics close to the saddles, for which more than one (typically many) exit options are available in their unstable manifold. These results suggest a positive role of noise for a range of natural and artificial information processing systems relying on complex state-space trajectories, as those may also rely on unstable states or similar structures, and may thus be of broad interdisciplinary interest.

\section{Networks of Oscillators and Heteroclinic Dynamics}
Heteroclinic networks naturally emerge in a variety of symmetrical systems composed of oscillators \cite{AT2005_2, Wordsworth2008, KGT2009, Neves2009}, both phase- and pulse-coupled, providing a model-independent framework for encoding and computing. We here consider networks of pulse-coupled leaky integrate-and-fire neurons. This simple model already captures the fundamental aspects required for the emergence of heteroclinic networks, i.e. symmetrical and excitatory delayed couplings. Furthermore, this model provides a closed-form solution for the system's time evolution between pulse events. This allows for efficient event-based simulations, when compared to more expensive time-based numerical integration.

Between pulse events, the dynamics of each node $i$ is defined by a voltage-like variable $V_i(t)$, satisfying the following differential equation and reset condition:
\begin{subequations}
  \begin{align}
  \frac{dV_i(t)}{dt} = -\gamma V_i(t) + A + W_i(t) + \Delta_i(t)  + \eta_i(t),\\
  \lim_{\zeta \rightarrow 0^+; \zeta > 0} V_i(t_f + \zeta) = V_{reset},
  \end{align}
  \label{eq:network}
\end{subequations}
where $\gamma$ is a dissipation parameter and $A$ a base driving current; the network coupling is given by ${W_i(t)=\sum_{\substack{j=1,\; j\ne i}}^N \epsilon \delta(t - \tau - t_{j})}$, a sum of incoming pulses at time $t$, where $\epsilon$ is the connection strength and $\tau$ the connection delay; $\Delta_i$ represents an external input source such that $\vert \Delta_i \vert \ll \vert A \vert$; and $\eta_i$ represents a Gaussian noise source. The second equation defines the reset condition that depends on the firing times $t_f$ of pulse events, which are themselves defined in terms of a threshold criterion $\{t_f  : V(t_f) = V_{threshold}\}$. Herein, we consider $V_{reset} =0$ and $V_{threshold}=1$. To preserve the existence of closed-form solutions for the system's time-evolution between events, we approximate $\eta_i$ through a pair of high-frequency, low-amplitude pulse generators which, in practice, just add a large number of new events. Each oscillator in the network is connected to two independent noise sources, one producing excitatory pulses, the other one producing inhibitory ones, for a mean input current of $0$ (see Supplementary Material).
\begin{figure}[]
\begin{centering}
\includegraphics[width=8.5cm,angle=0]{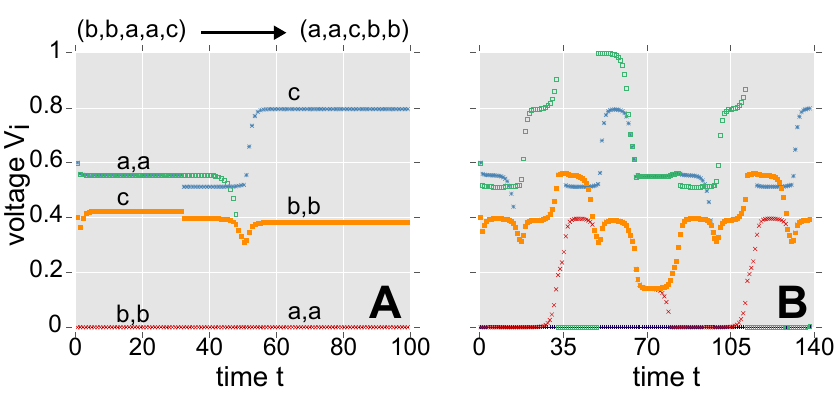}
\par\end{centering}
\caption[1st-entry]{ \small \textbf{Poincare sections for a single switch (A) and for a noise driven sequence of switches (B).} The potential $V$ of all 5 oscillators is plotted whenever $V_1 \rightarrow 0$, showing the cluster formation. Labels `a' indicates the unstable cluster, `b' the stable cluster and `c' the singleton. (\textbf{A}) at time 50, a single time perturbation induce a switch between two saddle states $(b,b,c,a,a) \rightarrow (a,a,c,b,b)$; (\textbf{B}) a low but persistent noise source induces a random sequence of visited states, resembling a random walk in a graph (the heteroclinic network itself).}
  \label{fig:dynamics}
\end{figure}

Saddle states arising in such networks of oscillatory neurons are characterized by the presence of poly-synchrony \cite{Neves2009}, that is, neurons synchronize in groups, here simply referred to as ``clusters''. While pulse-induced simultaneous resets promote synchronization, full synchronization is prevented by the delayed connections, because the neuron(s) sending the pulse(s) typically cannot synchronize with neurons receiving it(them). Some examples of poly-synchronous states were reported by Ashwin and Borresen\cite{Ashwin2004encoding} with permutation symmetry $S_3 \times S_2$, by Wordsworth and Ashwin\cite{Wordsworth2008} with permutation symmetry $S_2 \times S_2 \times S_1$ and by Neves and Timme\cite{Neves2012}  with permutation symmetry $S_{21} \times S_{21} \times S_{21} \times S_{21} \times S_{16}$. Here $S_n$ stand for all permutations of $n$ elements, i.e. all $n$ elements in a $S_n$ cluster are in an identical state. The $\times$ symbol is used to describe (compose) a state with multiple clusters. Furthermore, poly-synchronous saddle states often show instability to perturbation only over a single cluster. In this case, with the exception of one cluster, oscillators in each cluster are synchronized by incoming pulses causing their simultaneous reset, which also erases any small variation in voltage and thus the effect of any small perturbation. Neurons on the remaining cluster, here loosely called the ``unstable cluster'', are not reset by pulses, but rather independently reach threshold. Small voltage differences in the unstable cluster actually keep increasing at each cycle (one pulse per neuron) due to the fact that, the greater the neuron's voltage, the quicker an incoming pulse will cause it to reach to its threshold, in an exponential fashion. Lastly, networks of states exhibiting persistent switching dynamics are typically formed by states with the same symmetry. That is, all states in the network are simple permutations of each other, thus exhibiting the same stability properties after permutation.

\begin{figure}[]
\begin{centering}
\includegraphics[width=8.5cm,angle=0]{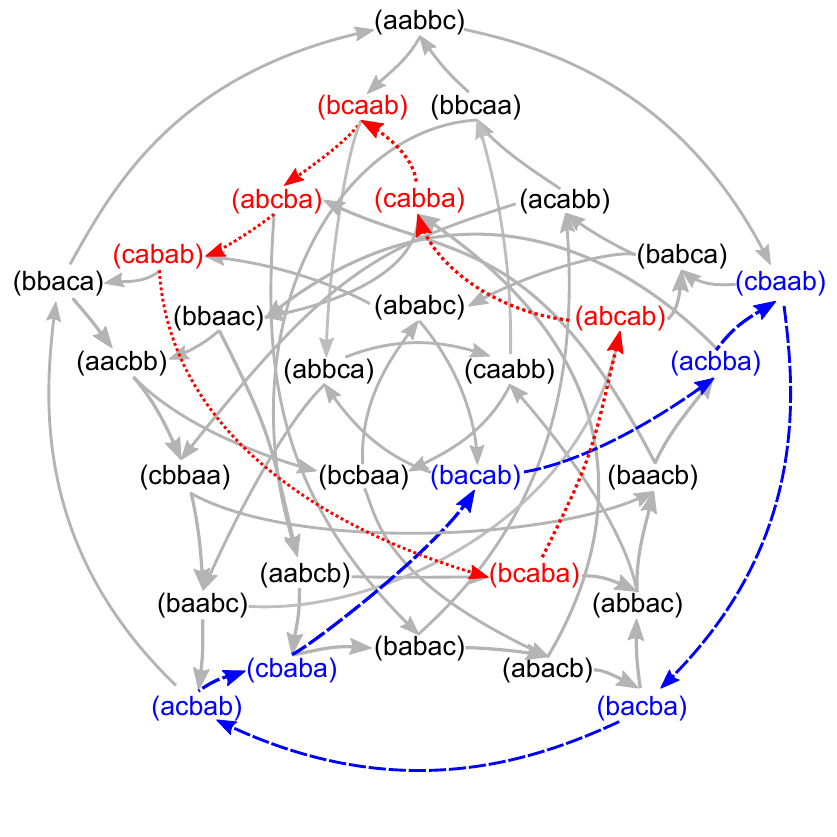}
\par\end{centering}
\caption[1st-entry]{ \small \textbf{Graph representation of the heteroclinic network defined by Equation~\ref{eq:rule} and all its permutations.} Each node represent a saddle periodic orbit; each arrow represent one heteroclinic orbit connecting two saddles. Two periodic sequences of states (complex trajectories) encoding the same input $\Delta=(\Delta_1,\Delta_2,\Delta_3,\Delta_4,\Delta_5)$ are highlighted with a red-dotted and blue-dashed pattern.}
  \label{fig:heteroclinic_network}
\end{figure}

As the calculation of the Mutual Information (in the next section) requires averaging the results of many trials over all orderings of inputs and initial conditions, and both these quantities grow exponentially with the system's size, we here choose a small but representative network of five neurons as a concrete example. For parameters $\epsilon = 0.025$, $A = 1.04$, $\tau = 0.49 \cdot \ln\bigl(A / (A-1)\bigr)$, the system exhibits saddle orbits with a three-cluster formation and $S_2 \times S_2 \times S_1$ permutation symmetry (see Figure~\ref{fig:dynamics}), i.e. two groups of two synchronized oscillators and one singleton. Moreover, it has been established before \cite{Neves2012, Ashwin2004encoding} that these states are unstable only to perturbation to one of their clusters. Let $\{a,b,c\}$ be labels for the clusters and $\{a\}$ be the label for the unstable cluster. We denote each saddle with a symbolic vector, e.g. $(a,a,b,b,c)$, in which each component corresponds to one of the $N=5$ oscillators. Such a vector uniquely labels a saddle and explicitly denotes the permutation symmetry of these states, as we can simply permute the vector to express any other state. By doing so, we obtain a total of ${5\choose{2,2}}=30$ saddle states. These states are interconnected via heteroclinic orbits following a simple transition rule \cite{Neves2009}: given a general perturbation $\Delta=(\Delta_1,\Delta_2,\Delta_3,\Delta_4,\Delta_5)$ where $\Delta_{1} > \Delta_{2}$, then
\begin{equation}
(a,a,b,b,c) \rightarrow (c,b,a,a,b),
\label{eq:rule}
\end{equation}
where the arrow denotes the dynamical switch between two saddles via a heteroclinic connection (see Figure~\ref{fig:dynamics}). In other words, the oscillator in the unstable cluster receiving the largest perturbation component becomes the new singleton, the original stable cluster loses stability and becomes the new unstable cluster, and the remaining two oscillators synchronize forming a new stable cluster (see Figure~\ref{fig:dynamics}a for an example). By permuting this relation, we obtain the set of all heteroclinic connections forming the heteroclinic network. Notice that because all nodes have the same characteristics and the connections are symmetric, heteroclinic networks can be represented as directed graphs \cite{Ashwin2010Designing}, in which the saddles are the nodes and the edges are the connections, see Figure~\ref{fig:heteroclinic_network}.

\begin{figure}[]
\begin{centering}
\includegraphics[width=7.5cm,angle=0]{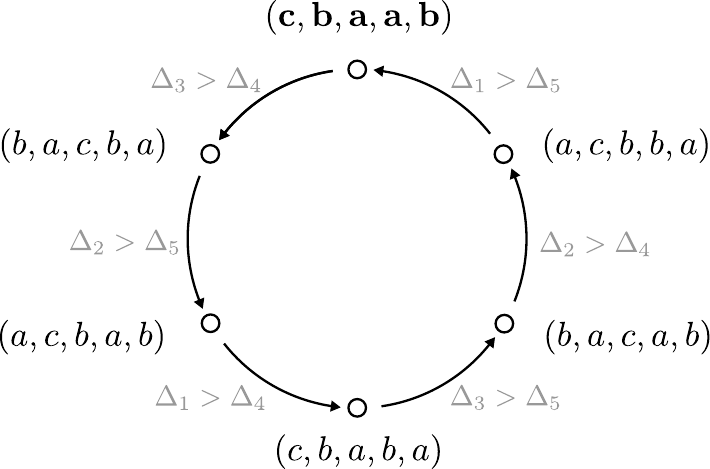}
\par\end{centering}
\caption[1st-entry]{\small \textbf{Detailed representation of a cyclic sequence of 6 saddle orbits, blue-dashed trajectory in Figure~\ref{fig:heteroclinic_network}}. The sequence was generated by an input with $\Delta_1 > \Delta_2 > \Delta_3 > \Delta_4 > \Delta_5$, starting from the  $(c,b,a,a,b)$ state. The relative ordering of the input components, here shown in gray, drives the transition between states. Conversely, observing a transition in the noiseless system at each saddle in the sequence implies an ordering between the two input components over the unstable cluster (marked with $a$ in the symbolic vector labeling the saddle state). In this example, observing the complete sequence reveals that $\{\Delta_1,\Delta_2,\Delta_3\}$ are all larger than $\{ \Delta_4,\Delta_5\}$.}
\label{fig:periodic_orbit}
\end{figure}

In this work, we are interested in characterizing the transmission of information through heteroclinic networks. It is thus of particular interest to understand how long-lasting signals are processed. In has been shown \cite{Neves2012} that, in the absence of noise, for every input having the same partial ordering, the sequence of approached saddle states deterministically realizes one of two possible trajectories approaching 6 specific saddles, depending on initial conditions. For example, given an input signals $\Delta_1 > \Delta_2 > \Delta_3 > \Delta_4 > \Delta_5$ and an initial state $(c,b,a,a,b)$, the system realizes the orbit shown in Figure~\ref{fig:periodic_orbit}. Permuting any two neurons between the $S_2$ clusters, yields the other possible orbit. 
In the noiseless case, we ignore any potential transient towards this orbit, because once it is approached, the system is locked there for as long as the signal is present. Statistically speaking, the transient doesn't play any relevant role. Given the transition rule in Equation~\ref{eq:rule} and all its permutations, observing the periodic orbit in Figure~\ref{fig:periodic_orbit} reveals that $\{\Delta_1,\Delta_2,\Delta_3\}$ are all larger than $\{ \Delta_4,\Delta_5\}$. Generalizing this result for all possible orbits, shows that the system computes a $k=3$-winners-take-all function over $N$ inputs, i.e. it determines the three strongest input signal components. Note that no additional information about the overall rank order is known from observing a cyclic sequence of saddle orbits.

The introduction of noise, which would be present in any realization of the system in the physical world, changes the input-driven switching dynamics in at least  one fundamental way. Whereas the noiseless system's dynamics is characterized by the approach of six saddle orbits, noise introduces an element of randomness, disrupting the cycle. For the rest of this work, we conceptualize our system as an input-output device, receiving an input and producing state sequences of some arbitrary length (see Figure~\ref{fig:saddle_sequence}c) as an output. Furthermore, because the state-to-state switching is driven by the difference between currents rather the values themselves (see Supplementary Material for more details), in what follows we will define the input current vectors in terms of the difference $D_i =\Delta_{i+1} - \Delta_{i}$ between its components.

\section{Quantifying Information in noisy Heteroclinic communication channels}
\label{sec:numerics}

In what follows, we show how to quantify information transmission via heteroclinic networks, studying them as noisy communication channels. Specifically, we measure the Mutual Information Rate (MIR) between input signals and the resulting outputs, subject to different noise levels. Formally, the mutual information between two random variables (r.v.) $\{X,Y\}$, here respectively taking values from the set of possible inputs signals $\mathcal{X}$ and the set of possible output responses $\mathcal{Y}$, is defined as the difference between the marginal entropy of the input r.v. $X$ and the conditional entropy between the input r.v. $X$ and the output r.v. $Y$, that is
\begin{equation}
  I(X;Y) = H(X) - H(X \vert Y),
  \label{eq:MI}
\end{equation}
where the marginal entropy $H(X) = -\sum_{x}p(x) \log p(x)$ measures the uncertainty about the variable $X$, while the conditional entropy $H(X \vert Y) = -\sum_{y}p(y) \sum_x p(x \vert y) \log p(x \vert y)$ measures the uncertainty in $X$ given that $Y=y$ is known, averaged over all possible $y$'s. Here $p(x)$ is the probability of a signal $x$ being transmitted and $p(x\vert y)$ is the conditional probability of $x$ given that $y$ is known. Therefore, their difference $I(X;Y)$ measures a drop in uncertainty in $x$ by observing $y$. For example, if $y$ predicts $x$ with absolute certainty, $H(X | Y) = 0$ and $I(X;Y) = H(X)$. On the other hand, if $X$ and $Y$ are independent,  $H(X | Y) = H(X)$ and $I(X;Y) = 0$. We define the MIR simply as $\mu_sI(X;Y)$ where $\mu_s$ is the average number of saddle states visited per unit of time.
Calculating the mutual information between the input and output of a system from Equation~\ref{eq:MI} relies upon estimating the three distributions $p(X)$, $p(Y)$ and $p(X \vert Y)$. Notice that no analytical formulation is available for $p(X \vert Y)$ for the system studied in this paper. Thus, we approximate these distributions. To do so, we first properly define our input set $\mathcal{X}$ and output set $\mathcal{Y}$; choose our source of noise; and finally, numerically compute the probabilities.
\vskip 0.2in
\noindent \textit{\textbf{Input Set:}} In this work, each input $x \in \mathcal{X}$ to the system (the external input source in Equation~\ref{eq:network}) is simply a vector of small and constant currents $\Delta = (\Delta_{1},\Delta_{2}, \ldots, \Delta_{5})$ with $\Delta_{i} \in \Re$ targeting oscillator $i$.

As the computation performed by heteroclinic networks is that of a partial ordering of their input currents, and we want all orderings to be equally represented, the chosen input set $\mathcal{X}$ must contain vectors with elements in all possible orderings in the same proportion. Therefore, we first pick a random input $\Delta^g$ ($g$ stands for ''generating''); and we then generate our input set $\mathcal{X} = \mathcal{P}(\Delta^g)$, the set of all possible permutations of $\Delta^g$. The cardinality of the resulting input set is thus $\left| \mathcal{X} \right| = 5!$. Notice that the defining computation at the vicinity of each saddle is the direction of the pairwise differences between the input components, rather than their magnitude. Thus, to generate one instance of $\Delta^g$, we randomly generate the differences between consecutive pairs of inputs. Specifically, we generate vectors of the form
\begin{align}
  \Delta^g_i &=
             \begin{cases}
               b &i=1\\
               \Delta_{i-1} + D_{i-1} &i=2, \ldots, 5,
             \end{cases}
    \label{eq:rv_generating}
\end{align}
where $b \in \mathbb{R}$ is some constant, and $D_1, \ldots, D_4$ are independent and identically distributed random variables from a uniform distribution in the interval $(0, 10^{-5})$. Permuting each generated vector in all possible configurations provides the complete set of inputs for each instance of our simulations. Furthermore, to better characterize the system response to signals and noise, we simulate the system for a variety of randomly generated input sets.

\noindent \textit{\textbf{Output Set:}} To define the general form of the output $y \in \mathcal{Y}$ of heteroclinic networks, we describe a network of states as a directed graph $G = (V, C)$, in which vertices $V$ are the set of sufficiently close neighborhoods of saddle states $S$ and the edges $E$ are the heteroclinic connections ${C = \bigl\{c \bigm| c=(s_i, s_j), \;\> s_i, s_j \in S \bigr\}}$ between those states (see Figure~\ref{fig:saddle_sequence}c). Any sequence of states can be represented as a walk on this graph.
A set $\mathcal{Y}$ thus has the general form of the set of all walks of some finite length $n$ on $G$, that is
\begin{equation}
  \mathcal{Y} = \mathcal{O}_n = \bigl\{ w \bigm| w = (s_i)_{i=1}^{n}, \;\> s_i \in S, \;\> (s_i,s_{i+1}) \in C \bigr\},
  \label{eq:output}
\end{equation}
with $y$ being an element of this set. We remark that $n$ is a hyperparameter that will be chosen taking in consideration arguments of convergence of measured information and the numerical computability time.

\noindent \textit{\textbf{Quantifying noise:}} The effect of noise on the MIR between input and output is proportional to its strength compared to the input amplitude. For this reason, we introduce a quantity relating the strength of input and noise. Because the dynamical response of the system depends fundamentally on the differences $D_i$ between signals, which are drawn from a uniform distribution, we introduce a signal-to-noise ratio defined as follows:
\begin{equation}
  \text{SNR} = \frac{E[D^2]}{\text{Var}[Z]},
  \label{eq:SNR}
\end{equation}
where $D$ is a uniformly distributed r.v., $E[D^2]$ is the expected value of its square, and $\text{Var}[Z]$ is the variance of the noise r.v. $Z$ (see Supplementary Material). Stronger noise leads to a lower SNR, and vice-versa.

\noindent \textit{\textbf{Numerical simulations:}} As discussed above, our objective is to characterize a noisy heteroclinic information channel in terms of MIR. To do so, we numerically approximate the distribution $p(Y \vert X)$ of the channel's outputs from $\mathcal{Y}$ given inputs from $\mathcal{X}$, as they are defined above, under varying levels of noise. Furthermore, our specific choice for the length of our outputs (here a hyperparameter) is $11$, determining the output set $\mathcal{O}_{11}$, i.e. the set of all walks of length $n=11$ on $G$. The value $n=11$ has been chosen as a trade-off between clarity of presentation of the results and increase in computation time needed to analyze the data. Results for different choices of $n$ are reported in the Supplementary Material; we observe the same qualitative results for all $n \ge 7$.

\begin{figure}[]
\begin{centering}
\includegraphics[width=8.5cm,angle=0]{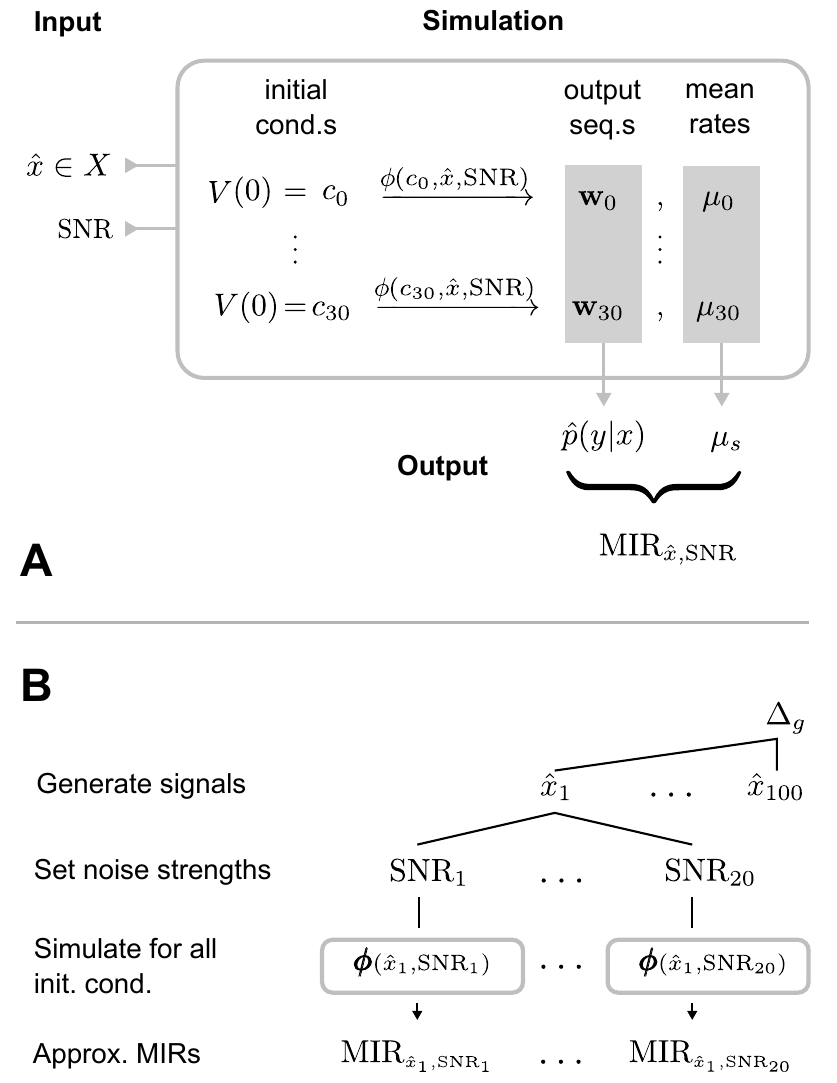}
\par\end{centering}
\caption[1st-entry]{\small \textbf{Layout of the numerical simulation suite used in this study.} (\textbf{A}) A single simulation takes a vector of currents $\hat{x}$ and a signal-to-noise ratio  $\text{SNR}$ level as an input, and internally simulates the system over 1000 state switches for all the 30 initial conditions (one run $\phi(c_l, \hat{x}, \text{SNR})$ for each clustered state $c_l$, with $l=1, \ldots, 30$, in the simulated heteroclinic network). For each initial condition, the simulation generates one set of output sequences $\pmb{w}_i = \bigl\{(s_j)_{j=t}^{t+n} \in \mathcal{O}_n \text{, with } t \in \{1,\ldots, 1000-n\} \text{, and } n=11 \bigr\}$ and one mean switching rate $\mu_i$. The aggregate set of output sequences is used to compute the approximated distribution $\hat{p}(Y \vert X)$, and the mean switching rates are averaged to an aggregate $\mu$; with these, it is possible to compute the mutual information rate associated to the given input set $\mathcal{X} = \mathcal{P}(\hat{x})$ (i.e the set of input vectors resulting from all possible permutations of the elements of $\hat{x}$) and noise strength as parameterized by the $\text{SNR}$. Due to the symmetries of the system, it is possible to derive a full $\hat{p}(Y \vert X)$ distribution from a partial $\hat{p}(Y \vert \hat{x})$ distribution obtained by simulating the system with a specific $\hat{x}$ example, as opposed to simulating the system for the full $X$ input set.
(\textbf{B}) The system is simulated for each of the $(\hat{x}_j, \text{SNR}_k)$ combinations of randomly generated input vectors $\hat{x}_j$ and $\text{SNR}_k$ levels, with $j \in 1, \ldots, 100$ and $k \in 1, \ldots, 20$, from which 2000 mutual information rates $\text{MIR}_{\hat{x}_j, \text{SNR}_k}$ are computed.}
\label{fig:simulations}
\end{figure}

As shown in Figure \ref{fig:simulations}, we generate $100$ MIR-SNR curves, allowing us to better understand the system's properties. We start by generating $100$ input sets $\mathcal{X}_i$ via $\mathcal{X}_i = \mathcal{P}(\Delta^{g}_i)$, the set of all permutations of $\Delta^{g}_i$, where $i$ indicates a specific set instance. For each input set $\mathcal{X}_i$ we pick only one element $\hat{x}_i$ to serve as an input for a simulation (see Figure \ref{fig:simulations}A). Each simulation is actually a set of numerical simulations using the same input, where we test $20$ different noise levels sampled logarithmically in the $[10^{-1}, 10^1]$ interval. Furthermore, we run the system starting from each of the $30$ possible states in the network, with independent noise realizations. Thus, a ``simulation" actually consists of $20 \times 30$ independent runs. For each level of noise and initial condition, we collect a switching sequence of length $k=1000$. To extract the needed $11$-walks from each complete switching sequence $q = (s_i)_{i=1}^k$, a moving window function $g_n(t) = (s_x)_{x=t}^{t+n}$ is used. For one simulation $(c=30, k=1000, n=11)$, the total number of collected output $y$ sequences for any element $\hat{x}$ of the input set and one SNR is equal to $29700$.

Note that a simulation using only one element $\hat{x}_i$ from $\mathcal{X}_i$ not only provides an approximation for $p(Y \vert \hat{x})$ but also an approximation for the full $p(Y \vert X)$ distribution. Due to the symmetries in the system's network of states and oscillator connectivity, any real $p(Y \vert x_{k})$ distribution for this system equals any other $p(Y \vert x_{j})$ up to a simple reordering of the vectors' elements. In this way, a single simulation run can actually provide an approximation of the full $p(Y \vert X)$ distribution. Then, by taking into consideration that we can marginalize $p(Y \vert X)$ over $X$ to obtain $p(Y)$, and that  $p(X)$ is a uniform distribution, where every input is equally likely, we obtain $p(X \vert Y) = \frac{p(Y \vert X)p(X)}{p(Y)}$ by Bayes' theorem. We thereby compute the mutual information as defined in Equation~\ref{eq:MI}. By multiplying this quantity by the average number of state switches per unit of time in the simulation run, we obtain the MIR of the run.

To summarize, we simulate the network for $100$ different inputs; for each, we test $20$ different noise levels; for each noise level, we start the system from all $30$ initial states. In this way we obtain $100$ curves of mutual information rate as a function of the SNR (again, see Figure \ref{fig:simulations} for an overview).

\begin{figure*}[t]
\begin{centering}
\includegraphics[width=17cm,angle=0]{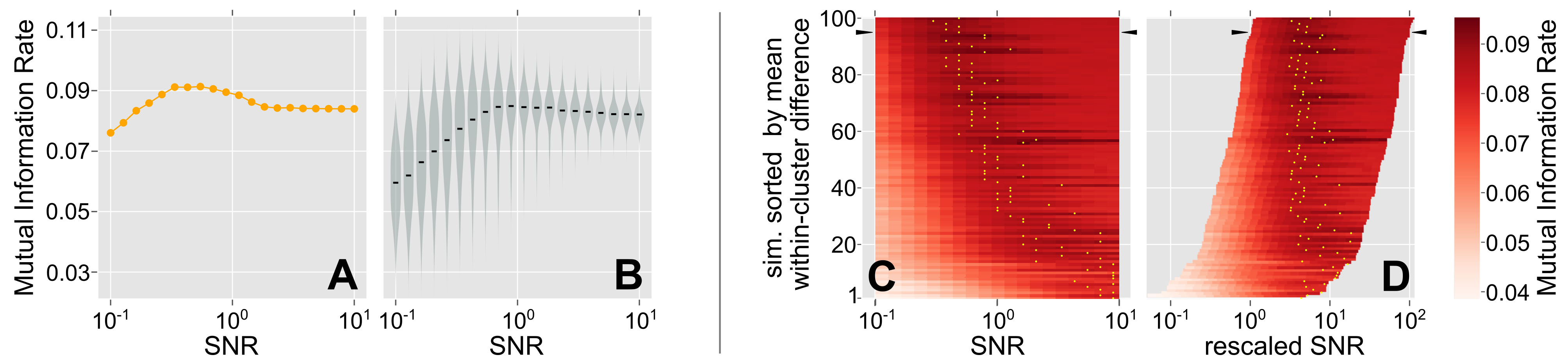}
\par\end{centering}
\caption{\small \textbf{The four panels in this figure report MIR-SNR curves as generated by the full numerical simulation suite depicted in Figure \ref{fig:simulations}.} (\textbf{A})
 MIR-SNR curve for a selected simulation. Here we report a single curve to highlight the presence of a characteristic increase in MIR for intermediate levels of noise, found for most of the simulated input sets. (\textbf{B}) Distribution (in grey) and median (in black) of the 100 MIR values computed for each SNR level from randomized input sets. The MIR distributions, shown rotated and mirrored for each SNR level, are estimated through Gaussian kernel density estimation. (\textbf{C}) MIR-SNR of 100 randomized input sets, ordered by decreasing mean difference between the three strongest and the two weakest input currents (a proxy measure for the ``strength'' of the cyclic sequence of 6 states). The maximum MIR for each simulation is highlighted by a yellow dot. (\textbf{D}) The same MIR-SNR of panel (\textbf{C}), where each simulation's SNR is rescaled by the mean difference between the strongest three currents and the two weakest.
 } 
 \label{fig:4_panels}
\end{figure*}

\section{Results}
Our computational experiments reveal that moderate levels of noise increase the MIR in the system in a predictable fashion (see Figure \ref{fig:4_panels}). Specifically, for intermediate levels of noise we observe an increase in MIR of up to 15\% with respect to the MIR measured at the smallest noise level tested ($\text{SNR}=10$).
\begin{figure*}[]
\begin{centering}
\includegraphics[width=17cm,angle=0]{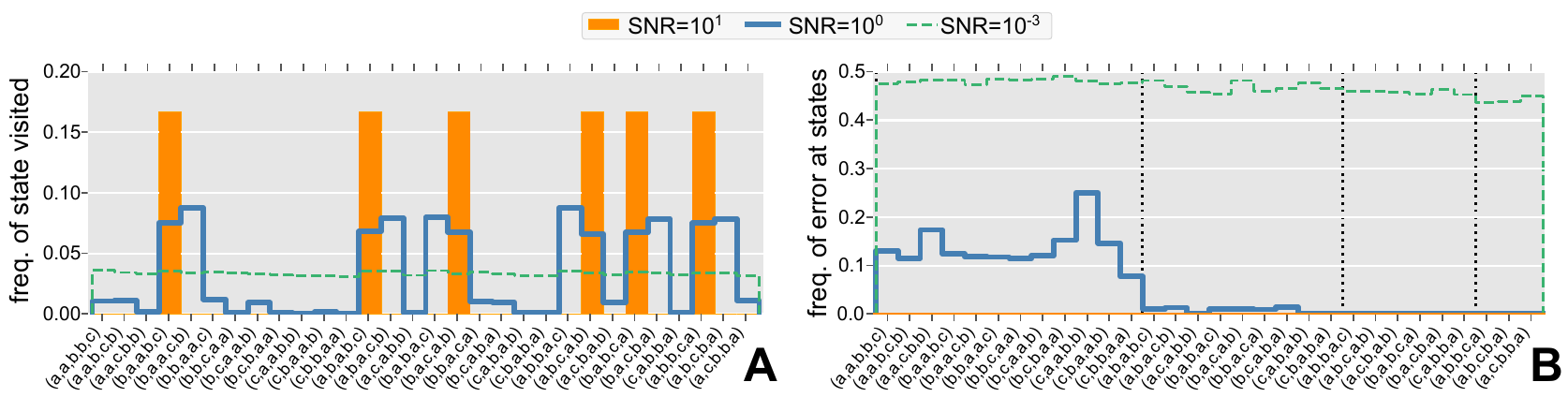}
\par\end{centering}
\caption{\small \textbf{Frequency of visits to each state (A) and frequency of wrong switches from each state (B).} Values are calculated from sequences of $n=1000$ states, for 3 levels of SNR, from an simulation run where the difference between consecutive input pairs $D_i = \Delta_{i+1} - \Delta_{i}$ is set to be $10^{-5}$ for all $i$. (\textbf{A}) When $\text{SNR}=10$, the switching traces the same orbit over and over again, and only 6 states are ever visited, no errors are made. When $\text{SNR}=1$, the noise causes a small amount of wrong switches, allowing the dynamics to explore a larger part of the network of states and move between two orbits representing the same input; thus, 12 states are visited most frequently. With $\text{SNR}=10^{-3}$, noise promotes a large amount of errors during switching and the switching process becomes virtually random. Panel (\textbf{B}) is divided in 4 sections by vertical dotted lines. The first section contains all those switches in which pair of inputs with the smallest difference $D_i = 10^{-5}$ are compared; the second section contains switches comparing inputs that differ by $2 D_i$; and so on for the third and fourth sections. The decreasing trend in switching error frequency for the four sections is caused by the increasing differences in input currents, which increases the effective SNR. For details on our numerical analysis, see Section \ref{sec:numerics}.}
\label{fig:trade_off}
\end{figure*}
This effect relies on the nature of the computation performed by heteroclinic networks: at each state, a different feature of the input is computed, i.e. which is the strongest input signal component over the unstable cluster, but only a subset of all states is ever visited in the noiseless system. Figure \ref{fig:trade_off} shows how these dynamics are modified by noise, by presenting how often each state is visited and the frequency of error for each state, for different levels of noise in runs of $n=1000$ recorded states. Here we say an error occurred during switching from state $A$ if the resulting state $B^*$ is different from the state $B$ predicted for a deterministic noiseless system.  
For low noise, a heteroclinic network only ever visits a subset of all states. In the specific case of the small leaky integrate-and-fire network we analyze, the switching dynamics are essentially confined to an orbit of six states, thus performing the same comparisons between pairs of input currents over and over again.
For high noise, the switching dynamics become highly unpredictable and largely independent of the input; this is reflected by the high spread of state frequencies in Figure \ref{fig:trade_off} and high error frequency for each state, approaching chance.
For intermediate noise, instead, occasional errors performed at some states allow the system to explore more of the network of states via short transient orbits, thus computing a greater range of input features (comparisons between pairs of input currents) and providing more information per time. The noise is low enough for mostly predictable orbits to exist, but high enough for the dynamics to be varied. That is, correctly computed transients orbit towards correct periodic orbits arise. 

Because each switch between saddles computes the largest input signal between a pair of inputs, there is a trade-off between the increased exploration due to noise, and a decreased accuracy of the computation at each switch: at intermediate noise, the system performs a more diverse range of computations on the input, although with lower accuracy at the level of the single computation. Approaching the system as a channel, shifts the focus from computation to information transmission and puts the richer information content to the forefront. Furthermore, higher noise levels are associated with a faster rate of switching (see Supplementary Material), thus it may also positively impact the MIR. Concurrently, the switching rate is also affected by the absolute strength of the difference between input currents, i.e. greater differences are associated to a higher rate of switching, but not to a larger error rate or a larger exploration. Together, these two features account for the shifting maximum in Figure \ref{fig:4_panels}, where inputs with smaller mean difference between currents tend to have their maximum MIR at higher SNRs (i.e. lower noise levels).
Note that the reason behind the ``disalignment'' in the plot curves in panel (\textbf{C}) of Figure \ref{fig:4_panels} is ultimately due to the formulation of SNR, which simplicity facilitates exposition, yet does not capture the full relationship between signal and noise in the system. In fact, rescaling each simulation's SNR by the mean difference between the three strongest currents and the two weakest as in Panel (\textbf{D}) of the Figure, substantially aligns the MIR results.

To put our results into perspective, we now shortly discuss how to interpret the reported increase in MIR. The fundamental constraint on our information measure is our choice of output. By choosing sequences of saddle states, we constrain the possible knowledge about the inputs to their full rank order (when observing the output), because each sequence of two saddles encode only the rank order between two input signals. In the noiseless case, in our example of dynamics only six comparisons are ever made and repeated cyclically, revealing a partial rank order. Particularly, an increase in MI (as reported) mathematically implies an increase in the amount of knowledge that can be gathered about the input by observing the output. Given our choice of encoding, an increase in MI can only mean that more about the full rank order is known. What exactly is learned depends on details of the system, the input, and the noise level (and type) and is therefore outside the scope of this article. Our results thus simply show, in a system agnostic way, that noise is capable of increasing the MI (knowledge about the complete rank order) and MIR in heteroclinic information channels, for this particular encoding. \\  

\noindent \textit{\textbf{Markov Chain analysis:}}
to show that our results really hinge on a simple trade-off between local errors in state-to-state transitions and global exploration of the network of states, and thus generalize beyond the specific choice of pulse-coupled system analyzed here, we now turn our attention to the heteroclinic networks' graphs. 
As previously discussed, a heteroclinic network can be described as a directed graph, and sequences of state transitions as walks on this graph. A persistent input signal induces cycles on the graph, with each transition corresponding to a comparison between input components, i.e. state-to-state switch. In the noiseless case, all switches are deterministic. When noise is present, transitions become probabilistic and the probability of "correct" switches (as prescribed in the noiseless case), decreases with increasing noise strength. 

Assuming that walks on this probabilistic graph exhibit the Markov property (i.e. the probability of the next state only depends on the current one), the graph can be seen as a Discrete-Time Markov chain, where a given input $x \in X$ and a comparison success probability vector $\vec{c}_{s_i}$ define the transition matrix $P_{s_i, s_j}$. For any such system, it is possible to derive the success probability vector $\vec{c}_{s_i}$ (or equivalently, the error probability vector $1 - p_c$) associated with a given noise level. Pulse-coupled systems such as the subject of our previous simulations, for example, do exhibit the Markov property, because of the memory-erasing effect of simultaneous pulse-driven resets in their stable clusters. The probability vectors $\vec{c}_{s_i}$ for three different SNR are shown in Figure \ref{fig:trade_off} (\textbf{B}).

For the sake of exposition, we here simplify analysis by setting a single parameter $p_c$ as a global comparison success probability at each state. Note that the $p_c$ probability parameter is tied to the noise level in the heteroclinic network the Markov chain is abstracting. Manipulating this parameter is akin to manipulating a signal-to-noise ratio (SNR) parameter in the system implementing the heteroclinic network, as we have done in our previous simulations. As the SNR decreases, the probability of a correct transition approaches chance. Similarly, in this analysis, we manipulate the $p_c$ parameter in the range between determinism $p_c = 1$ and chance $p_c = \frac{1}{n_t}$, with $n_t$ being the number of possible transitions at each state.

For $p_c < 1$, the Markov chain is ergodic, allowing for the analytical derivation of the limiting distribution $\psi$ of its states, i.e. the probability for each state $s$ to be the active state at time $t$, for $t \rightarrow \infty$. By knowing the limiting distribution $\psi$ and the transition matrix $P$, it is possible to calculate the probability of any $n$-walk $y = (s_i)_{i=1}^{n} $ as follows:

\begin{equation}
p\big( (s_i)_{i=1}^{n} \big) = \psi_{s_1} \prod_{i=2}^n P_{s_{i-1}, s_{i}}
\label{eq:walk_probability}
\end{equation}

It is then straightforward, given any input $x \in X$ and comparison success probability $p_c < 1$, to calculate the probability distribution $P(Y \vert x \in X)$ of output $n$-walks $y \in Y$ on the graph and, thus, to calculate the Mutual Information $I(X; Y)$ between inputs and walks. 

For $p_c = 1$, the Markov chain is non-ergodic, thus requiring a different approach for the calculation of walk probabilities and resulting Mutual Information. In this case, given an input, all state sequences converge to one of a finite number of cycles. In the limit of $t \rightarrow \infty$, only $n$-walks on those cycles have probability greater than zero, because any transient walks any of the cycles will only ever happen once, and the corresponding probabilities will thus converge to zero as all subsequent walks are confined to one of the cycles. The initial state, however, stills determines which specific cycle is approached. 
If we assume a uniform probability over starting states, it is possible to derive the probability of observing a cycle by simply taking the proportion of starting states eventually leading to that cycle. In turn, the probability of any given $n$-walk, is either $0$, if the $n$-walk is not a walk on one of the cycles, or equal to the probability of the traced cycle divided by the number of possible $n$-walks on that cycle. Having derived the $n$-walk probability distribution for a given input, it is thus possible to calculate the Mutual Information $I(X; Y)$ between inputs and walks.

\begin{figure}[]
\begin{centering}
\includegraphics[width=8.5cm,angle=0]{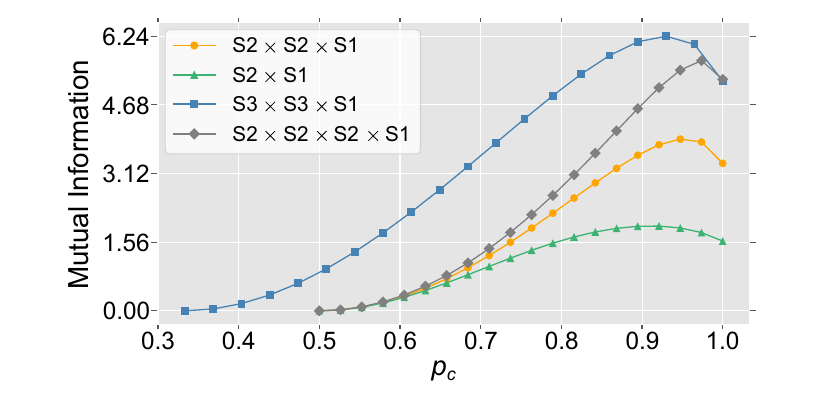}
\par\end{centering}
\caption{\small \textbf{MI curves resulting from a Markov chain analysis of known heteroclinic networks.} 
 MI curves for different probability of correct switches $p_c$ ($p_c = 1$ for the noiseless case) for $\text{S2} \times \text{S2} \times \text{S1}$, $\text{S2} \times \text{S1}$, $\text{S2} \times \text{S2} \times \text{S2} \times \text{S1}$, and $\text{S3} \times \text{S3} \times \text{S1}$, which have been reported on phase- or pulse- coupled networks or both. For all symmetries, one of the large clusters is unstable and the neuron receiving the strongest input becomes the new singleton. We again consider walks of length $11$. Note that for all system exhibiting a $\text{S2}$ cluster, chance is associated to $p_c = \frac{1}{2}$ (2 transitions per state), while for the system exhibiting $\text{S3}$ unstable clusters, chance is associated to $p_c = \frac{1}{3}$ (3 transitions per state).
 }
\label{fig:markov_panels}
\end{figure}

In Figure \ref{fig:markov_panels}, we show the result of this analysis performed on three graphs corresponding to four known heteroclinic networks. All of these display the same non-monotonicity emerging from the previously discussed numerical simulations. In particular, the peak MI emerging from the Markov chain analysis of the $\text{S2} \times \text{S2} \times \text{S1}$ system, closely resembles the one found in numerical simulation (see Supplementary Material). 
Note that, for this analysis, the number of $n$-walk distributions to be considered in order to compute the Mutual Information depends on the size of the input set. As this grows factorially with the number of oscillators in the heteroclinic system, systems with more than a few oscillators are still numerically challenging. For this reason, our analysis is here limited to smaller systems. 

\section{Conclusion}

In this work, we studied how noise and input signals jointly affect the Mutual Information Rate (MIR) in heteroclinic communication channels, shifting the focus from heteroclinic computing to information transmission. As a concrete example, we have focused our efforts on a system of delta-pulse-coupled oscillators, studying how the signal-to-noise ratio (SNR) controls the measured input-output MIR. Specifically, we studied how the magnitude of pairwise differences between input components interacts with noise to control the MIR. Interestingly, MIR is a non-monotonic function of the SNR: for small SNR the dynamics are dominated by noise-triggered, random state-switches, thus exhibiting the lowest MIR; for large SNR, the dynamics almost exclusively exhibit deterministic switches triggered by the signal, yielding a cyclic trajectory approaching a specific sequence of states and, thus, exhibiting a well-defined MIR (here taken as baseline); for a considerable range of intermediate values of SNR, the MIR increases from its baseline value, before falling to its minimum, thus exhibiting non-monotonicity. This occurs due to a trade-off between a small amount of noise-triggered ``wrong'' turns and a wider exploration of the network of states, where a wrong turn can trigger a new deterministic transient trajectory returning to a cyclic trajectory. 
The overall result is a larger variety of comparisons between the input components by approaching a larger variety of saddles, at the cost of a small amount of computing errors. From the point of view of information transmission, this translates to more knowledge being transmitted about the input signal overall, albeit a less certain one at each switch event, due to noise.

Our choice of network of oscillators was dictated by practical considerations of numerical simulation, due to the large amount of simulation trials required to accurately compute the MIR, and because many properties of heteroclinic networks, e.g. number of states, grow exponentially with the system size. Notwithstanding the specific implementation of heteroclinic network considered in this study, our results are general as they do not (qualitatively) depend on the system size, oscillator model or the specific heteroclinic network realization, but only on the existence of a heteroclinic network of unstable states and saddles' unstable manifolds with more than one direction. For any such system, there will be a trade-off between uncertainty at each state switch and a resulting greater exploration of the network of states, leading to MIR sweet spots for given SNR ranges.
To support this view, we presented a Markov chain analysis of networks of symmetrical saddle states and show that, from a system-agnostic perspective, different networks exhibit qualitatively the same non-monotonic MI curves. 

Our results may have direct implications on a variety of interdisciplinary issues concerning computation in natural and artificial systems. Notably, heteroclinic dynamics have been suggested as an underlying mechanism for the olfactory dynamics in animals \cite{laurent2001odor, huerta2004learning, afraimovich2004heteroclinic, AT2005}. In this context, our results on increased MIR through noise suggest yet a new role for noise in neural information processing and transmission, adding to works on: ``stochastic resonance'' \cite{wiesenfeld1995stochastic, magalhaes2011vibratory, zeng2000human}, where noise facilitates the detection of sub-threshold inputs; on ``system size resonance''\cite{Pikovsky2002System}, where the system size becomes an order parameter; on ``coherence resonance''\cite{Pikovsky1997Coherence} where noise induces a more coherent (better synchronized) state; and, more generally, on ``stochastic facilitation'' \cite{mcdonnell2011benefits}. For artificial systems, our results reveal a clear picture of how noise affects computation and information transmission in heteroclinic networks composed of symmetrical states (under index permutation) and potentially on systems exhibiting complex state-space trajectories, e.g hierarchical heteroclinic networks \cite{Voit2018Hierarchical}, networks of saddle-states composed of states with different symmetries \cite{Stone1999Noise, Armbruster2003Noisy, Bakhtin2010Small, Ashwin2010Designing} or models of specific features of the mind \cite{Afraimovich2004Origin,Afraimovich2018Mind,Rabinovich2018Discrete}, because their dynamics also rely on unstable states or similar structures. Notably, our numerical results are calculated on spiking neural networks with full resets, which promote an explicit loss of memory, i.e. simultaneous resets induced by incoming spikes instantly erase voltage differences between oscillators in the same stable cluster. Thus, even though our Markov chain analysis suggests a degree of generality, whether our results generalize to systems of phase-coupled units, where memory may fade exponentially fast, or not is still an open question. 

Overall, our results comes in a timely manner, as the microprocessor industry is exploring the use of ``imprecise'' processors to compute and transmit information with greater speed and lower power consumption \cite{palem2005energy, chakrapani2007probabilistic}.

\section*{Supplementary Material}
\label{sec:SupMat}
\begin{figure}[!h]
\begin{centering}
\includegraphics[width=8.5cm,angle=0]{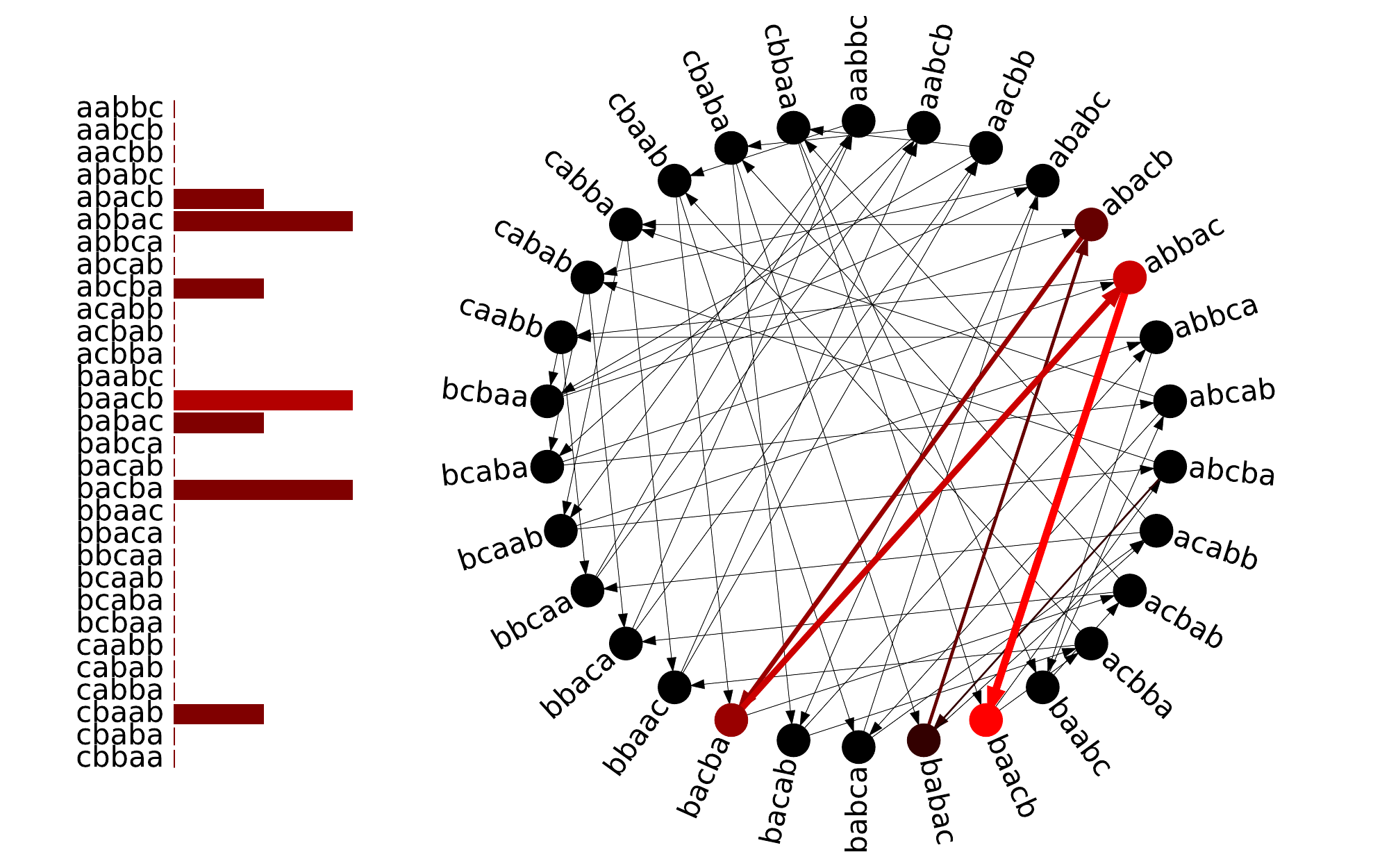}
\par\end{centering}
\caption{ \small
\textbf{Sequence of visited states}. The associated video demonstrate sequences of visited states and their relative frequency of occurrence for different signal-to-noise ratios. (Multimedia view)}
\label{fig:deltaxprob}
\end{figure}

In the supplementary manuscript we explain in more detail how we implemented noise in our system and how it affects its dynamics, such that the results can be reproduced. Furthermore, we present the results on the mutual information and mutual information rate calculated over different path lengths. Finally, to provide some intuition on how we generate the statistics in our work, we also provide a video, see Figure~\ref{fig:deltaxprob} (Multimedia view), showing transitions between states for short intervals of time for different signal-to-noise ratios. 

\section*{Acknowledgements}
\label{sec:thanks}
We would like to acknowledge Ikerbasque (The Basque Foundation for Science) and moreover, this research is supported by the Basque Government through the BERC 2018-2021 program and by the Spanish State Research Agency through BCAM Severo Ochoa excellence accreditation SEV-2017-0718 and through project RTI2018-093860-B-C21 funded by (AEI/FEDER, UE) and acronym ``MathNEURO''. Partially supported by the German Research Foundation (Deutsche Forschungsgemeinschaft, DFG) under project number 419424741 and under Germany's Excellence Strategy -- EXC-2068 -- 390729961 -- Cluster of Excellence Physics of Life at TU Dresden, and the Saxonian State Ministry for Science, Culture and Tourism under grant number 100400118.

\section*{Data Availability}
The data that supports the findings of this study are available within the article [and its supplementary material].

\label{sec:data}

\bibliography{bibliography}

\end{document}


\title{Supplementary material:\\ Stochastic facilitation in heteroclinic communication channels}
\maketitle

\tableofcontents

\section{Introduction}
\label{sec:orgheadline1}
In this supplementary material we added some results and explanations that are not essential to the main text but support it in a variety of ways. We explain in more detail how we implemented noise in our system and how it affects its dynamics, such that the results can be reproduced. Furthermore, we present here the results on the mutual information rate calculated over different path lengths. Finally, to provide some intuition on how we generate the statistics in our work, we also added one video showing transitions between states for short intervals of time for different signal-to-noise ratios. 

\section{Noise implementation and switching times}
\label{sec:orgheadline6}
As noise plays a fundamental role on systems exhibiting unstable attractors, we explain in detail its implementation in our model. We also present a short review on the fundamental results on switching times in heteroclinic networks and how noise influences this measure.

\subsection{Quantifying noise}
\label{sec:orgheadline2}

\begin{figure}[t]
\begin{centering}
\includegraphics[width=8.5cm,angle=0]{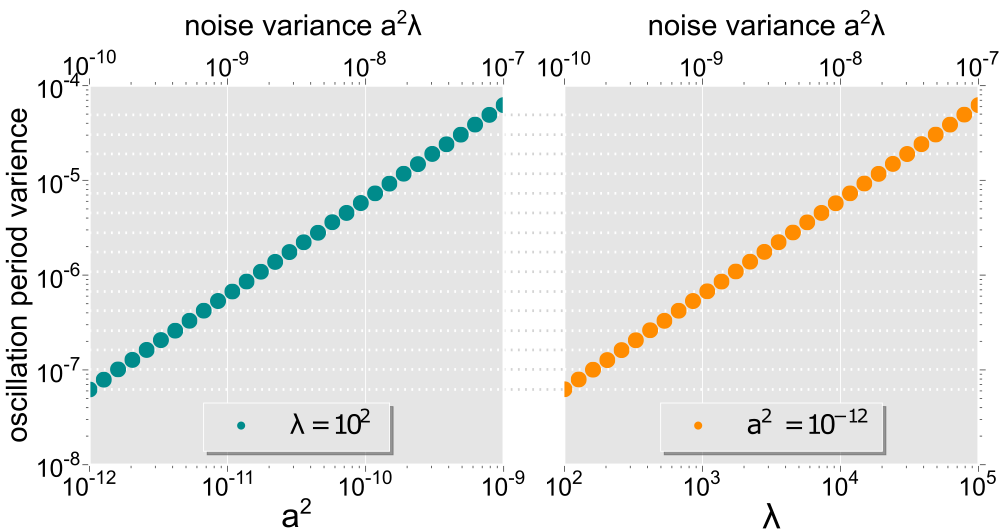}
\par\end{centering}
\caption[1st-entry]{\small \textbf{Reset timings variance for a range of noise parameter values.} On the left, a noise rate \(\lambda=10^2\) is fixed, and the squared noise amplitude \(a^2\) is manipulated. On the right, a squared noise amplitude \(a^2=10^{-12}\) is fixed, and the noise rate \(\lambda\) is manipulated. Note how, for the same resulting noise variance in the manipulation of the two parameters (shown in the top axis), the resulting reset timings variance in the left and right plot is approximately the same (highlighted by the dotted horizontal lines). \label{fig:lambda_vs_amp}}
\label{fig:lambda_vs_amp}
\end{figure}

In this work, noise is produced as high-frequency low-amplitude random pulses independently generated and applied to each oscillator in the network as external pulses. Two pulse-generators are in place for each oscillator, one producing excitatory pulses with rate $\frac{\lambda}{2}$ and amplitude $a$ and the second producing inhibitory pulses with rate $\frac{\lambda}{2}$ and amplitude $-a$. To avoid synchronization caused by noise-pulses, pulses are not produced in fixed time intervals, but follow a Poisson distribution. This is in contrast to fixed times intervals and varied amplitudes approach, that may lead to synchronization.

If we denote by $X_1$ and $X_2$ the number of pulses generated in a unit of time, respectively, by the excitatory pulse generator and the inhibitory pulse generator, their current contribution in a unit interval are equal to \(Y^+ = aX_1\) and \(Y^- = -aX_2\). These are, by definition, random variables. The mean and variance of \(Y^+\) and \(Y^-\) are given by,
\begin{align}
E(Y^+) &= E(aX_1) = aE(X_1) = a \frac{\lambda}{2} \\
E(Y^-) &= E(-aX_2) = -aE(X_2) = -a \frac{\lambda}{2} \\
V(Y^+) &= V(aX_1) = a^2V(X_1) = a^2 \frac{\lambda}{2} \\
V(Y^-) &= V(-aX_2) = a^2V(X_2) = a^2 \frac{\lambda}{2}
\end{align}
The overall current produced by both generators in a unit of time is itself a random variable which we denote with \(Z = Y^+ + Y^-\). As \(Y^+\) and \(Y^-\) are independent, the mean and variance of Z are equal to
\begin{align}
E(Z) &= E(Y^+ + Y^-) = E(Y^+) + E(Y^-) = 
a \lambda/2 - a \lambda/2 = 0\\
V(Z) &= V(Y^+ + Y^-) = V(Y^+) + V(Y^-) = 
2 \cdot a^2 \lambda/2= a^2\lambda 
\end{align}
For sufficiently high \(\lambda\) rate and sufficiently small amplitude \(a\), the produced pulse train approximates a continuous current, so that the only parameter controlling its effect on the system is \(V(Z)\). To verify that the noise works as intended, we ran simulations measuring the average period of a single (uncoupled) oscillator for different noise levels. The set the noise level we choose different combinations of \(\lambda\) and \(a\) values, as shown in Figure \ref{fig:lambda_vs_amp}. For each $\lambda$ and \(a\) values, the average period of the oscillator and its variance is calculated over \(100,000\) resets (oscillations).

Figure~\ref{fig:lambda_vs_amp} shows that our approximation of a continuous noise is indeed sufficient. Similarly to the continuous version of the noise (analytical), how strongly the noise affects the oscillator seem to only depend on the noise variance $a^2\lambda$ and not strongly depend on the frequency $\lambda$. This is fortunate, given that we chose to use an event-based simulator, the computation time in our simulations increases with the number of events to simulate. Increasing the spiking rate \(\lambda\) thus has a computational cost. There's no such cost associated to the manipulation of \(a\). For our simulations, we set $\lambda$ to $100$ and vary $a$ to achieve different noise levels.

\subsection{Switching times for a noiseless system}
\label{sec:orgheadline3}
In the absence of noise or any source of asymmetry, e.g. input signal, the dynamics of delta-pulse-coupled systems exhibiting heteroclinic networks will always precisely converge and reach an unstable attractor, from any point of its basin of attraction. This is in contrast to phase-coupled systems that would generate an ever slower orbit approaching the actual heteroclinic orbits.  

\begin{figure}[]
\begin{centering}
\includegraphics[width=8.5cm,angle=0]{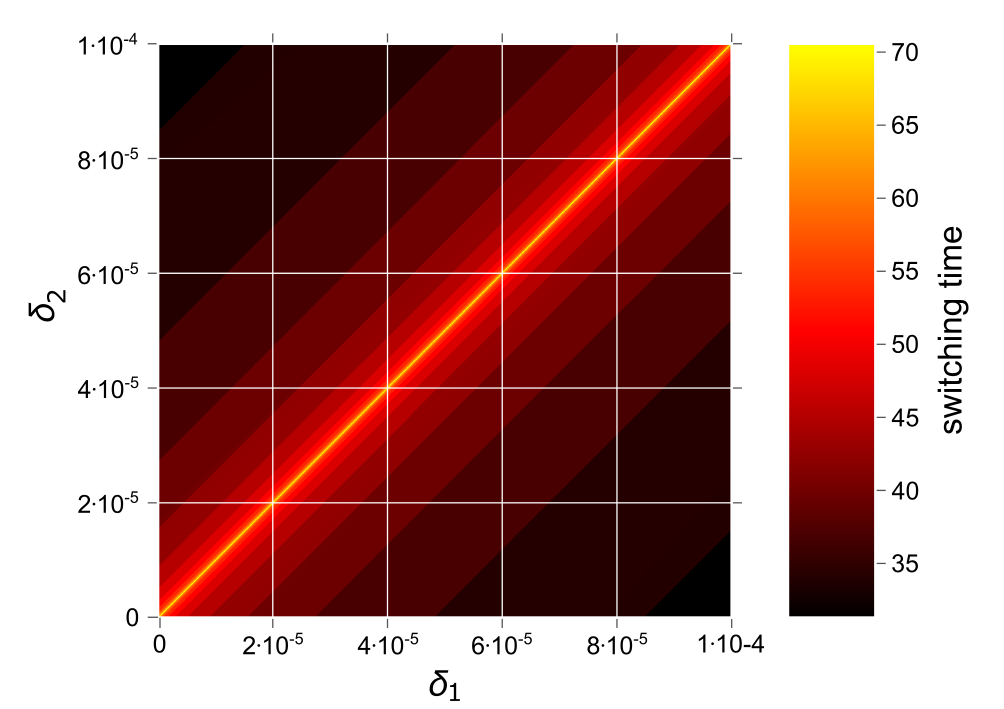}
\par\end{centering}
\caption{\small
\textbf{Switching time as a function of delta currents applied to oscillators in unstable cluster.} The figure shows how the switching time depends primarily on the difference between the symmetry-breaking currents to the unstable cluster, rather than their magnitude. }
\label{fig:switching_time}
\end{figure}

When signals are introduced, thus breaking the symmetry, the switching times between states are primarily determined by the magnitude of the difference between each pair of input components $(\delta_i, \delta_j)$ in the direction of the two unstable manifolds (see Figure \ref{fig:switching_time}). The remaining input components do not affect switching times, because they are completely erased by simultaneous reset events in all oscillators belonging to the stable clusters.

\begin{figure}[]
\begin{centering}
\includegraphics[width=8.5cm,angle=0]{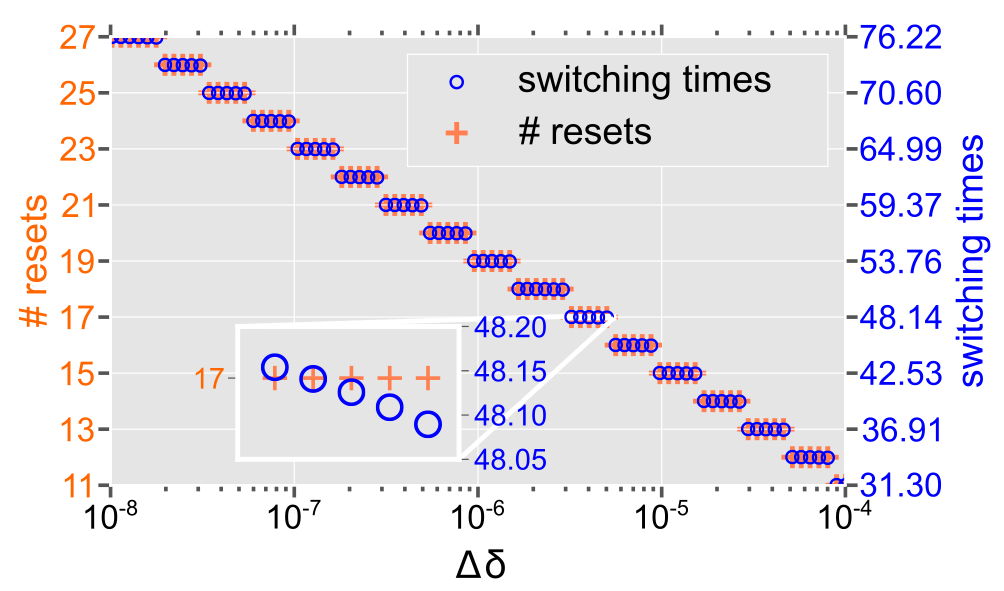}
\par\end{centering}
\caption{\small
\textbf{Switching time and number of resets as functions of delta difference}. This Figure highlights the exponential relationship between the difference in delta currents ( \(\Delta_{ij} = \delta_i - \delta_j, \delta_i > \delta_j\) ) to the unstable cluster and the switching time. In particular, notice how the discrete time jumps are related to the underlying change of the number of resets in switching.}
\label{fig:time_and_resets}
\end{figure}

As shown in Figure~\ref{fig:switching_time} as symmetrical plateaus and in more detail in Figure~\ref{fig:time_and_resets}, changes in switching times are characterized by discrete jumps. This is because the magnitude of the $\Delta_{ij} =\delta_i - \delta_j$ differences governs the number of resets needed between the de-synchronization and re-synchronization of the oscillators in a new clustered state. In this way, a smooth increase/decrease in the magnitude of the difference $\Delta_{ij}$ can lead to the addition/subtraction of a reset until re-synchronization, which entails a discrete jump in switching time. The differences in switching times inside each ``step'' is caused by faster switching times induced by the larger input difference. We remark that the switching time increases roughly exponentially as the magnitude of $\Delta_{ij}$ decreases, see Figure~\ref{fig:time_and_resets}.

\subsection{Switching times for a system with noise}
\label{sec:orgheadline4}
As shown in a variety of works \cite{neves2010universal, bakhtin2011noisy, ashwin2016quantifying,neves2017noise}, noise reduces the switching times between states in heteroclinic networks. The increased speed of switching is due to the fact that noise changes the distribution of times at different distances from the attractors: less time is spent near the attractor when noise is present. As the dynamics is slower near attractors and faster away from them, noise has the effect of accelerating the switching. In the range of noise strengths that allows switching dynamics to exist, i.e. does not break away from the heteroclinic network, the acceleration is proportional to the strength of the noise.

\begin{figure}[]
\begin{centering}
\includegraphics[width=8.5cm,angle=0]{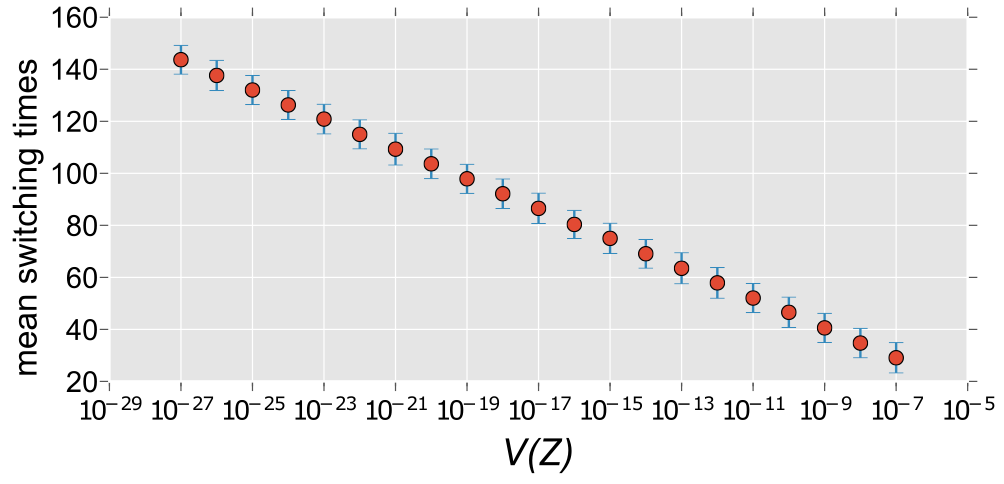}
\par\end{centering}
\caption{\small
\textbf{Mean and standard deviation of switching time for a range of values of noise variance.} The blue bar represents one standard deviation above and below the mean.}
\label{fig:mean_switching_time}
\end{figure}

In Figure~\ref{fig:mean_switching_time} we show the mean switching times and their standard deviations for different noise variances $V(Z)$ and no input present. Note that with no input, SNR can't be used to quantify the noise level, so we use the noise variance instead. The data was generated by simulating the system with no input for $20$ values of noise variance, recirding the switching times for sequences of $1000$ states.
As Figure \ref{fig:mean_switching_time} shows, increasing the noise variance accelerates the switching in the system roughly in a logarithmic manner. 

As shown recently in detail \cite{neves2017noise}, the combination of noise and inputs can be described as a combination of its limit cases. While for large input differences $\Delta_{ij}$ at low noise the switching times can be well approached by the noiseless case, the noise provides an upper boundary for switching times given comparatively small $\Delta_{ij}$. The intermediary cases are a combination, where the contribution of noise always accelerates the switching times for a fixed $\Delta_{ij}$ value.

\subsection{Errors in state-switching}
\label{sec:orgheadline5}

In an heteroclinic network, noise does not only affect the switching times between saddles, but also may change its computational properties. In the absence of noise, input signals promote deterministic state-to-state switches. For our $N=5$ network example, switches are determined by the $(a,a,b,b,c)\rightarrow(c,b,a,a,b)$ switching rule and all its permutations, see the main paper for a detailed description.
When noise is introduced, switches become stochastic and, thus, errors in the expected switching sequence can occur. That is, state-to-state transitions which don't respect the transition rules.

\begin{figure}[]
\begin{centering}
\includegraphics[width=8.5cm,angle=0]{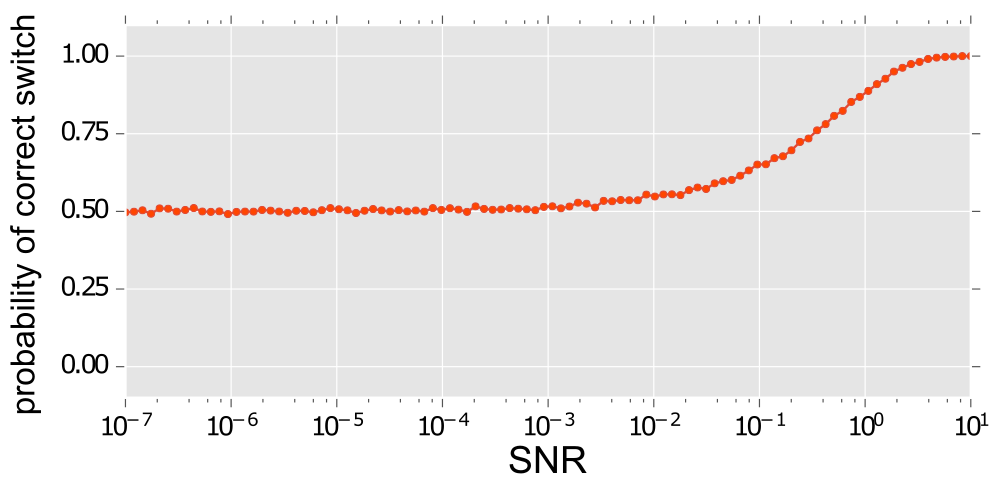}
\par\end{centering}
\caption{ \small
\textbf{Probability of correct switch for a range of SNR levels}. Notice how low SNR (i.e. high noise) leads to a chance level of correctly reaching the state determined by the transition rule in the noiseless system.}
\label{fig:deltaxprob}
\end{figure}

In Figure \ref{fig:deltaxprob} we show the results of a simulation where we fixed the input to have a single non-zero component coinciding with the unstable manifold, and vary the SNR by varying the magnitude of this component (but maintaining the noise parameters fixed). For each level of SNR, we run \(10^4\) trials by setting a starting state as initial conditions, and checking which state was reached at the end of the transition. In this way we are able to approximate the probability of reaching the right state for each of \(100\) levels of SNR from \(10^{-7}\) to \(10^{1}\).

As expected, at small SNR the probability of switching to the right state gets closer to chance levels; for large SNR, the switchings are deterministic; and the the probability of switching to the right state increases monotonically with increasing SNR.

\section{Sequence of visited states}
Here we introduce a supplementary video to our main manuscript, available in the web, showing three different sequences of states given three different signal-to-noise ratios (SNR) and concurrently displaying the comparative frequency in which each saddle is visited. A snapshot is provided below.
\begin{figure}[!h]
\begin{centering}
\includegraphics[width=8.5cm,angle=0]{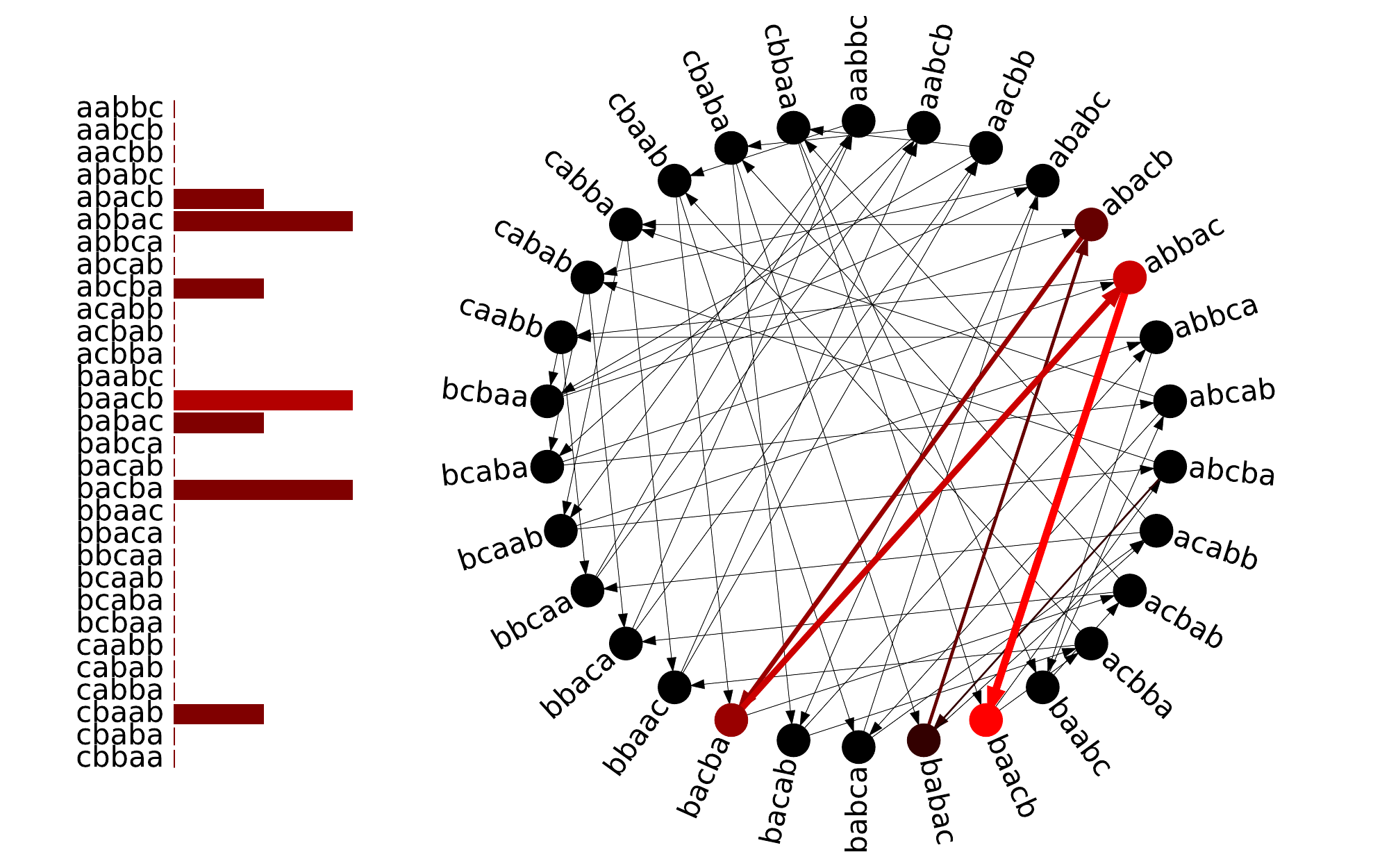}
\par\end{centering}
\caption{ \small
\textbf{Sequence of visited states}. The associated video demonstrate sequences of visited states and their relative frequency of occurrence.}
\label{fig:deltaxprob}
\end{figure}
To watch the video please look for the associated mp4 file.

\section{Results for varying walk lengths $n$}
\label{sec:orgheadline8}
\begin{figure*}[t]
\begin{centering}
\includegraphics[width=15cm,angle=0]{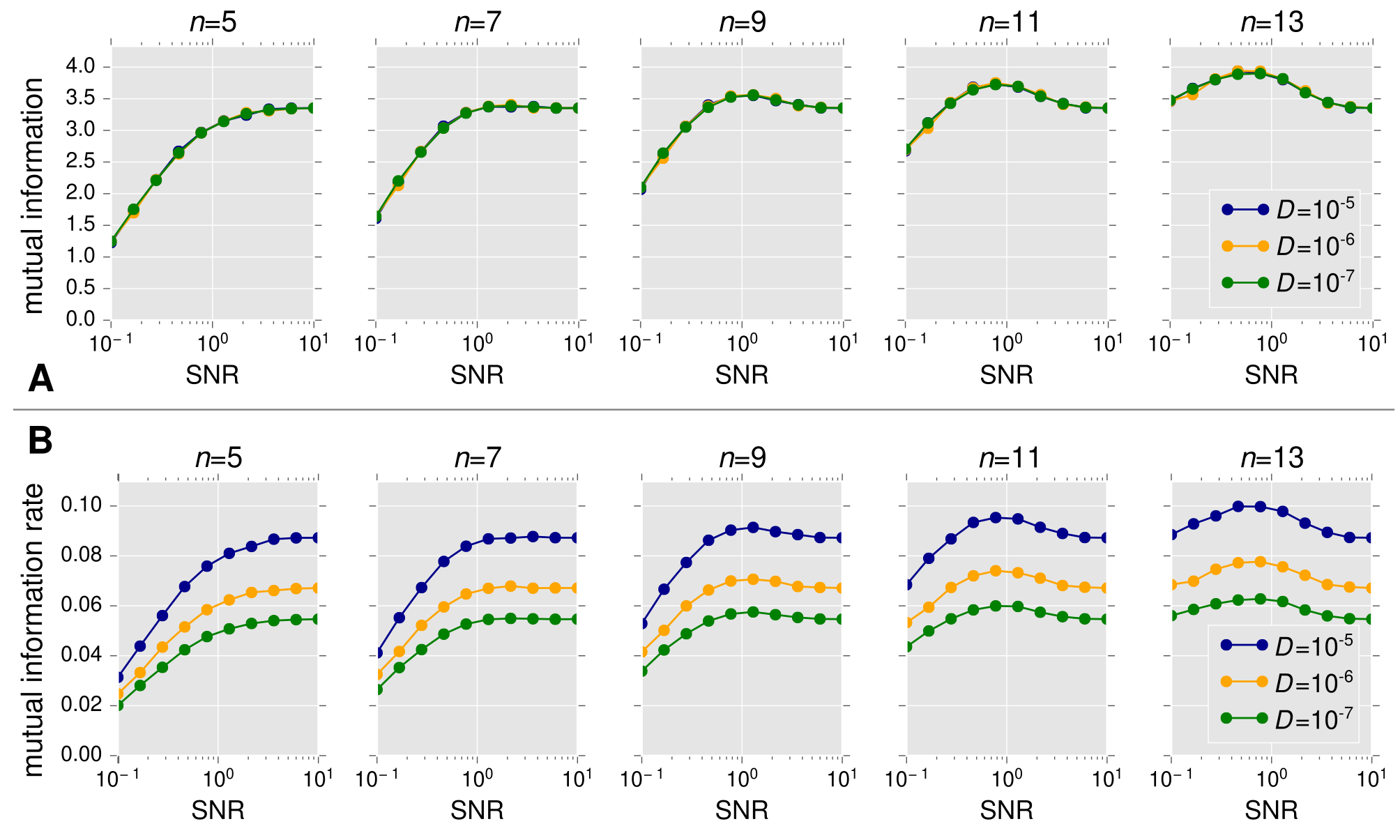}
\end{centering}
\caption{\small
\textbf{Mutual information and mutual information Rate for a range of output sequence's lengths.} \textbf{a} The results shown in this figure are qualitatively consistent with those presented in the main paper. Note that, for each plot, SNR increases moving left to right on the $x$ axis. Equivalently, moving right to left, noise levels get progressively stronger, and the dynamics of the system become less and less deterministic. \textbf{b} As for panel \textbf{a}, the results shown in this panel are also qualitatively consistent with those presented in the main paper. Again note that, for each plot, SNR increases towards the right on the $x$ axis. Equivalently, moving towards the left, the level of noise increases with respect to the signal, leading to less deterministic dynamics, which consequences are discussed in the main paper.}
\label{fig:mi_rate_results_n}
\end{figure*}

As discussed in the main paper, for intermediate SNRs, noise leads to an increased but controlled exploration of the underlying network of states.
Observing a switching between two specific states carries specific information about the input which drove the transition. When noise is present, observing a single switching carries less information than in the noiseless case, due to the added uncertainty on the ultimate cause for the switching (input versus noise). However, because of the increased exploration of the network for intermediate SNRs, a wider repertoire of switches can be observed and, thus, for longer sequences the loss of information at each specific switching is outweighed by the gain in information from the wider "coverage" in observable switchings.
Here we report the results of additional simulations, equivalent to the first set of simulations presented in the main paper, but performed with different choices of output sets. In particular, we here manipulate the length $n$ of the state sequences considered as output of the system. 

For each input set we test 20 levels of SNR, sampling logarithmically in the \([10^{-1}, 10^1]\) interval, and starting the simulations from each of the 30 network states. The output sets are here taken to contain all walks of length \(n=5, 7, 9, 11, 13\) on $G$. As shown in Figure 7, for $n \leq 6$ the output sequences are too short, such that the loss of information at each specific switching predominates. However, the effect becomes visible for $n > 6$, as sequences become long enough to capture the information made available by the increased exploration of the network of states. Note that the size of the output set doubles for each increase of $n$ (as each state is connected to other two states in the network, such that two sequences of length $n$ follow from each sequence of length $n-1$), and thus appropriately  approximating the output distribution requires exponentially more resources for simulation. This is why we have set $n=11$ as a trade-off between simulation time and clarity of results for the main paper, and $n=13$ as an upper limit for further exposition in the supplementary material. 

\bibliography{bibliography}